\documentclass[aps,pre,preprint,superscriptaddress,endfloats,showpacs]{revtex4}

\usepackage[utf8]{inputenc}
\usepackage{amsmath}
\usepackage{graphicx}
\usepackage{epstopdf}
\usepackage{color}

\newcommand{\erw}[1]{\langle #1 \rangle}

\begin{document}

\title{Theoretical study of electronic damage in single particle imaging experiments at XFELs for pulse durations 0.1 - 10 fs}
\date{\today}

\author{O. Yu. ~Gorobtsov}
	\affiliation{Deutsches Elektronen-Synchrotron DESY, Notkestra{\ss}e 85, D-22607 Hamburg, Germany}
	\affiliation{National Research Centre "Kurchatov Institute", Akademika Kurchatova pl., 1, 123182 Moscow, Russia}
\author{U.~Lorenz}
	\affiliation{Deutsches Elektronen-Synchrotron DESY, Notkestra{\ss}e 85, D-22607 Hamburg, Germany}
	\affiliation{Institute of Chemistry, University of Potsdam,  D-14476 Potsdam, Germany}
\author{N.M.~Kabachnik}
	\affiliation{Skobeltsyn Institute of Nuclear Physics, Lomonosov Moscow State University, 119991 Moscow, Russia}
	\affiliation{European XFEL GmbH, Albert-Einstein-Ring 19, D-22761 Hamburg, Germany}
\author{I.A.~Vartanyants}
\email[Corresponding author: ]{ivan.vartaniants@desy.de}
\affiliation{Deutsches Elektronen-Synchrotron DESY, Notkestra{\ss}e 85, D-22607 Hamburg, Germany}
\affiliation{National Research Nuclear University MEPhI (Moscow Engineering Physics Institute), Kashirskoe shosse 31, 115409 Moscow, Russia }

\begin{abstract}
X-ray free-electron lasers (XFELs) may allow to employ the single particle imaging (SPI) method to determine the structure of macromolecules that do not form stable crystals.
Ultrashort pulses of 10 fs and less allow to outrun complete disintegration by Coulomb explosion and minimize radiation damage due to nuclear motion, but electronic damage is still present.
The major contribution to the electronic damage comes from the plasma generated in the sample that is strongly dependent on the amount of Auger ionization.
Since the Auger process has a characteristic time scale on the order of femtoseconds, one may expect that its contribution will be significantly reduced for attosecond pulses.
Here, we study the effect of electronic damage on the SPI at pulse durations from 0.1 fs to 10 fs and in a large range of XFEL fluences to determine optimal conditions for imaging of biological samples.
We analyzed the contribution of different electronic excitation processes and found that at fluences higher than $10^{13}$-$10^{15}$ photons/$\mu$m$^2$ (depending on the photon energy and pulse duration) the diffracted signal saturates and does not increase further.
A significant gain in the signal is obtained by reducing the pulse duration from 10 fs to 1 fs.
Pulses below 1 fs duration do not give a significant gain in the scattering signal in comparison with 1 fs pulses.
We also study the limits imposed on SPI by Compton scattering.

\end{abstract}


\maketitle

\section{Introduction}

Modern x-ray crystallography methods make it possible to determine the structure of crystals at atomic resolution \cite{crystallography1, crystallography2}.
In the age of molecular biology, one of the most important questions in life science is the determination of protein structures.
Unfortunately, many protein macromolecules, especially membrane proteins, either do not crystallize or form only extremely small crystals \cite{Carpenter2008}.
Imaging such nanocrystals, or even individual macromolecules, is impossible with conventional x-ray sources, since the sample is destroyed before a high resolution diffraction pattern can be obtained \cite{Damage}.

These difficulties can be circumvented with the use of x-ray free-electron lasers (XFELs) \cite{LCLS,SCSS,XFEL}.
The ultrashort, intense, and coherent XFEL pulses can be used to obtain the diffraction pattern of a small, sub-micron sample before it is destroyed by a Coulomb explosion \cite{Neutze00}.
Conventional radiation dose limits \cite{Damage} are surpassed by orders of magnitude, since the dominant damage mechanism, the breaking of chemical bonds, can be outrun by short XFEL pulses.
In addition, for reproducible samples, many single shot patterns with random sample orientations can be taken and oriented to produce a full three-dimensional (3D) diffraction pattern \cite{orientation1,orientation2,orientation3,Yefanov}.
Using phase retrieval algorithms \cite{PhaseRetrieval1,PhaseRetrieval2}, a 3D image of the samples electron density can be reconstructed.
This approach is conventionally called nowadays single particle imaging (SPI) \cite{GaffneyChapman}.

In spite of significant progress in imaging biological samples at XFELs \cite{Seibert2011, Kimura2014, Barke2015, VanderSchot2015}, experiments have highlighted severe challenges for such single-particle imaging experiments.
Particles tend to show high conformational heterogeneity \cite{Maia2009,Hantke2014}, and the coherent scattering signal to background ratio is low.
To improve the signal level it has been suggested to use even more powerful pulses \cite{Hantke2014}.


However, the intensity of the x-ray pulse cannot be increased without limit.
At a certain threshold, extremely intense x-ray pulses strip all electrons from the atoms, leaving a sample that does not scatter any more.
In addition, statistical fluctuations in the random ionization of the atoms due to the quantum mechanical nature of these processes produce a background signal that dominates the diffraction pattern for strong ionization.
In our previous publication \cite{Ulf}, we have shown that x-ray induced electronic damage limits the pulse fluence that can be reasonably employed in an experiment.

It was also realized \cite{Hau-Riege04,Ulf} that a major ionization mechanism in SPI experiments is a secondary ionization by trapped Auger electrons.
At the same time, Auger decay has a lifetime of several femtoseconds, which points to the possibility to outrun the impact ionization by trapped electrons with extremely short, possibly attosecond XFEL pulses.
It was recently suggested that such pulses can be produced in principle \cite{Saldin2006, ShortTanaka}.

The aim of this paper is twofold.
On one hand, we study to what extend the use of ultrashort XFEL pulses reduces the electronic damage of a typical biological sample.
In particular we are interested in pulse durations from 0.1 fs to 10 fs, since this is the range where the suppression of the Auger decay is expected.
On the other hand, we analyse an extension of the model used in \cite{Ulf} by including additional ionization processes such as shake-off and Compton scattering.
We also calculate the contribution of Compton scattering to elastically scattered radiation measured on the detector (see also recent work \cite{SantraCompton} where Compton scattering from a carbon cluster was analyzed).

\section{Theory}

\subsection{Elastic scattering}
\label{sec::theory::scattering}

In the following we recall the description of a single particle coherent diffraction experiment in the frame of kinematical approximation (see for details Ref. \cite{Ulf} and Appendix \ref{appendix::scattering}).

In a typical experiment a large number of single-shot diffraction patterns at different orientation of the particles will be measured. In the following we will assume that all these diffraction patterns can be perfectly aligned and averaged. If the incident beam is fully coherent over the sample area and has a uniform intensity distribution, we obtain the following expression for the averaged scattered intensity \cite{Ulf}
\begin{equation}
<I({\bf q})>  =  \sum^N_{i,k=1} e^{-i{ \bf q}  \cdot ({\bf R}_k - {\bf R}_i)} \int dt J(t) <f_i^{*}({\bf q}, t) f_k({\bf q}, t)> \ .
\label{SII.1}
\end{equation}
Here, $\bf q$ is the scattering vector, $J(t)$ is the intensity of the incoming pulse, and $f_i({\bf q}, t)$, ${\bf R}_i$ are the time-dependent form factor and position vector of the \textit{i}-th atom, respectively. The brackets $\langle \ldots \rangle$ denote averaging over many pulses, and the summation is performed over all atoms in the sample.

As it was shown in our previous work \cite{Ulf} due to stochastic nature of the electronic damage process the expression (\ref{SII.1}) can be written as a sum of two terms
\begin{gather}
\label{theory::averaging2}
	\erw{I(\mathbf{q})} = I_{\text{W}}(\mathbf{q}) + I_{\text{B}}(q) \ ,
\end{gather}
where
\begin{equation}
            I_{\text{W}}(\mathbf{q}) = I_0 \sum_{i,j} W_{ij}(q) \exp [i\bf{q}(\bf{R}_j-\bf{R}_i)]
\label{SII.3}
\end{equation}
is the coherent signal containing the structural information and
\begin{equation}
            I_{\text{B}}(q) = I_0 \sum_i B_i(q)
\label{SII.4}
\end{equation}
is an incoherent background without structural information. Here $I_0 = \int J(t) \mathrm{d}t$ is the total fluence of the x-ray pulse.
The matrix $W_{ij}(q)$ and vector $B_i(q)$ are defined by the time-dependent average values $\erw{f_i(q, t)}$ and pulse to pulse fluctuations $\delta f_i(q, t)$ of the form factor $f_i(q, t) = \erw{f_i(q,t)} + \delta f_i(q,t)$ of each individual atom through the following relations
\begin{eqnarray}
  W_{ij}(q) & =&  \frac{1}{I_0} \int J(t) \erw{f_i^{\ast}(q, t)} \erw{f_j(q,t)} \mathrm{d}t  \ , \\
  B_i(q) & = & \frac{1}{I_0} \int J(t) \Big\langle |\delta f_i(q,t)|^2 \Big\rangle \mathrm{d}t  \ ,
\label{eq::background_element}
\end{eqnarray}
where spherically symmetric form factors were considered.
The structural term $I_{\text{W}}(\mathbf{q})$ (\ref{SII.3}) determines the degradation of the diffraction pattern due to evolution of the form factors while the photoionization process and the background term $I_{\text{B}}(q)$ (\ref{SII.4}) adds an additional background that is due to fluctuations of the individual form factors during the same process.

An additional background contribution comes from inelastic (Compton) scattering.
The Compton signal at the detector is given by
\begin{gather}
\label{theory::compton_intensity}
	I_\text{Compton}(q) = \sum_i \int J(t) \erw{S_i(q,t)} dt,
\end{gather}
where $\erw{S_i(q, t)}$ is an averaged incoherent scattering function of the atom $i$ (see for details Appendix \ref{appendix::rates}) and brackets have the same meaning as before.
Equations \eqref{theory::averaging2} - \eqref{theory::compton_intensity} were used in our simulations of diffraction patterns from a biological sample.

\subsection{Rate equation implementation}
\label{sec::theory::rates}

To determine the time-dependent average form factors $\erw{f_i(q,t)}$, their fluctuations $|\delta f_i(q,t)|^2$, and the average incoherent scattering function $\erw{S_i(q,t)}$ for each constituent atom type $i$, a rate equation approach \cite{Sang-Kil} was implemented.
First, we define a set of states that the atom can potentially occupy.
As such states, we consider the electronic ground states for all possible occupations of the electronic shells.
For example, the carbon atom can have between zero and two electrons in each of the $1s$, $2s$, and $2p$ shells, yielding a total of 27 states.
The time-dependent occupation probabilities $p_{\xi; i}(t)$ for the $\xi$-th state
of the atom were obtained by solving a set of coupled differential equations
\begin{gather}
    \label{appendixB::rate_equation}
    \dot p_{\xi; i}(t) = \sum_{\eta\neq \xi} R_{\xi\eta; i}(t) p_{\eta; i}(t)
    - R_{\eta\xi; i}(t) p_{\xi; i}(t)
    \ .
\end{gather}
Here, $R_{\xi\eta; i}(t)$ denotes the total time-dependent rate of transition from state $\eta$ to state $\xi$ for atom type $i$.
We assume that initially all atoms are in the ground state.
The solution of the differential equations (\ref{appendixB::rate_equation}) yields the time-dependent occupation probabilities $p_{\xi; i}(t)$ for each state $\xi$ of the specific atom.

The form factors $f_{\xi; i}(q)$ for each state $\xi$ were obtained from electronic wave functions calculated within the Hartree-Fock-Slater (HFS)
approximation \cite{HermanSkillman}.
Within this model, the average form factors $\erw{f_i(q,t)}$ and their fluctuations $\erw{|\delta f_i(q,t)|}$ are given by
\begin{align}
\label{appendixA::formfactor1}
    \erw{f_i(q,t)} &= \sum_{\xi} p_{\xi;i}(t) f_{\xi;i}(q) \,,\\
\label{appendixA::formfactor2}
    \erw{|\delta f_i(q,t)|^2} &= \erw{|f_i(q,t)|^2} - |\erw{f_i(q,t)}|^2 =
    \sum_{\xi} p_{\xi;i}(t) |f_{\xi;i}(q)|^2 - |\erw{f_i(q,t)}|^2
    \ .
\end{align}
Note that in Eqs.~\eqref{appendixA::formfactor1}-\eqref{appendixA::formfactor2}
we used explicit state-dependent form factors without the additional
assumption that they scale with the number of bound electrons as in Ref. \cite{Hau-Riege07}.
In the frame of our approach, the valence shells contract significantly on ionization of the core electrons, and, as a consequence, the corresponding form factors expand in reciprocal space (see Ref. \cite{Ulf}).

The average time-dependent inelastic scattering function $\erw{S_i(q, t)}$ can be calculated in a similar way \cite{Hubbell01}
\begin{gather}
\label{appendixB::incoherent_formfactor}
	\erw{S_i(q,t)} = \sum_\xi p_{\xi;i}(t) S_{\xi;i}(q,t)
	\,
\end{gather}
where $S_{\xi;i}(q,t)$ is the inelastic scattering function for the state $\xi$
\begin{gather}
	S_{\xi; i}(q) = Z_{\xi;i} - \sum_{r=1}^{Z_{\xi;i}}|f^{r}_{i}(q)|^2
	\ .
\end{gather}
Here $Z_{\xi;i}$ is the total number of electrons in the atom in state $\xi$, and $f^{r}_{i}(q)$ is the form factor of the $r$-th electron in the atom.
Here we neglect effects that may forbid excitation of an electron from one orbit to another due to Pauli exclusion principle.

The time-dependent total transition matrix $\hat{\mathbf{R}}(t)$ contains contributions from several electronic processes

\begin{gather}\label{eq:matrix}
    \hat{\mathbf{R}}(t) = \hat{\mathbf{R}}^{\text{photo}}(t) + \hat{\mathbf{R}}^{\text{Auger}}
    + \hat{\mathbf{R}}^{\text{shake}}(t) + \hat{\mathbf{R}}^{\text{Compton}}(t)
    + \hat{\mathbf{R}}^{\text{escape}}(t) + \hat{\mathbf{R}}^{\text{trap}}(t)
    \ ,
\end{gather}
where $\hat{\mathbf{R}}^{\text{photo}}(t)$ is the rate of direct photoionization, $\hat{\mathbf{R}}^{\text{Auger}}$ is the Auger decay rate, $\hat{\mathbf{R}}^{\text{escape}}(t)$ and $\hat{\mathbf{R}}^{\text{trap}}(t)$ are the rates of secondary ionization produced by escaping and trapped electrons, respectively.
These four terms have been considered in our previous work \cite{Ulf}.
Here we also take two additional ionization channels into account, namely shake-off processes with the rate
$\hat{\mathbf{R}}^{\text{shake}}(t)$ and ionization due to Compton scattering with the rate $\hat{\mathbf{R}}^{\text{Compton}}(t)$, which can be important
 at high x-ray energies.
Notice that the first four terms are purely atomic, while the latter two are collective effects.
See appendix~\ref{appendix::rates} for details on the evaluation of the rates.

In the model of electronic transitions used in this paper we assume that the electron plasma thermalizes instantaneously, i.e., the thermalization process is much shorter than the pulse duration.
This is considered as a good approximation for comparatively long pulses \cite{Hau-Riege04}.
For x-ray pulses as short as $100$ as it is necessary to investigate this question in more details.
It is well established that non-homogeneous trapped electron gas is formed on very early stages of x-ray pulse particle interaction \cite{Hau-Riege04, Hau-Riege12}.
To estimate these relaxation times the following arguments are typically used.
Electrons emitted from a center of the spherically symmetric particle of radius $R$ are trapped if their kinetic energy is lower than $E_{trap}^{0}=e^2 R^2 n/3\epsilon_0$, where $e$ is the electron charge, $n$ is the charge number density and $\epsilon_0$ is the permittivity of vacuum.
For Auger energies about $E^{Auger} \sim 250$ eV for the particle with the radius of $R=15$ nm we obtain the charge number density
$n \sim 2 \cdot 10^{-4}$ ${\AA}^{-3}$ at which Auger electrons are trapped by the ionized particle.
Such small charge density corresponds roughly to $3 \cdot 10^{-3}$ electrons being ionized per atom.
Assuming that the dominant ionization process is the direct photoionization for the flat-top x-ray pulse we get an estimate for the charge number density
$n(t) \sim \sigma^{photo} n_{at} F  (t-t_0)/T$, where $n_{at}$ is atomic density, $F$ is the pulse fluence, and $t_0$, $T$ are times of the pulse start and duration, respectively.
From this relation we obtain that the trapping time scales inversely with the photoelectron cross section and fluence and is proportional to pulse duration.
Our estimates show that for all pulses below $1$ fs and fluences considered in this paper the formation times of non-stationary trapped electron plasma are below $10$ as.

At the same time, thermalization process of this non-homogeneous, trapped electron gas takes place on much longer time scales.
Calculations of characteristic thermalization times performed according to \cite{Spitzer} give an estimate of about few femtoseconds.
By extending our model to shorter pulse durations we slightly overestimate the ionization rate of trapped electrons.
However, this contribution at pulse durations below $1$ fs is already significantly lower than the contribution from other ionization processes.
This is due to the fact that for very short pulses Auger electrons do not contribute to ionization process, while secondary ionization by escaping and shake-off electrons are producing only low energy secondary electrons that can not effectively ionize.
By these arguments we can extend our model to times as short as $100$ as, keeping in mind that we still slightly overestimate the contribution from secondary ionization.

In our model, we assume that the lowest non-vanishing order perturbation theory (LOPT) is valid for high energy x-rays in the range of powers up to $10^{26}$ W/cm\textsuperscript{2} and for pulse durations down to $100$ as.
This assumption is based on the fact that ionization for high photon energies is well described in the frame of LOPT, if the pulse duration is significantly larger than the field period (see e. g. Ref. \cite{Lambropoulos}).

\section{Results and discussion}
\label{sec::results}

\begin{figure}
	\includegraphics[width=\linewidth]{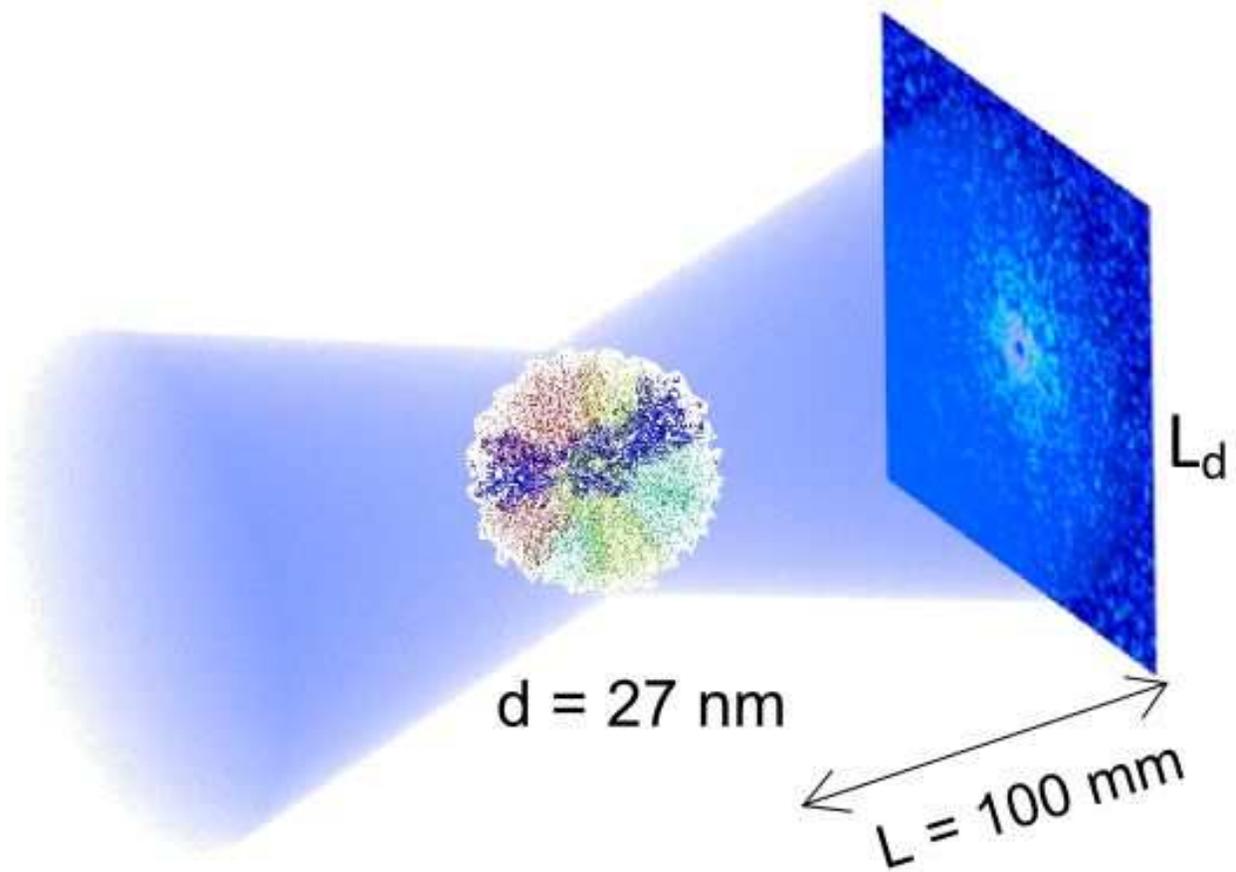}
	\caption{\label{fig::experiment}
		(Color online) Schematic of a single particle coherent X-ray diffractive imaging experiment.
		A single FEL pulse illuminates the sample of size $d$ from the left and scatters from it, with the diffraction pattern measured by a detector of size $L_d$ at a distance $L$ from the sample.
	}
	\newpage
\end{figure}

To analyze the effect of electronic radiation damage, we simulated SPI experiment as sketched in Fig.~\ref{fig::experiment}.
For the sample, we used a human adenovirus penton base chimera shell \cite{sample}.
It has a dodecahedral shape with a diameter of 27 nm and contains about 200 000 nonhydrogen atoms, giving a mass density of about 0.5 g/cm$^3$.
To account for typical virus densities we have increased this mass density value by three in ionization calculations.
The ratio between carbon, oxygen and nitrogen atoms in this sample is approximately 3:1:1.
In our simulations, we neglected the contribution from hydrogen and sulfur atoms to the ionization dynamics and the scattering.

We performed simulations for photon energies of 3.1 keV and 12.4 keV with the experimental parameters listed in Table~\ref{tab::parameters}.
The lower energy is experimentally attractive due to higher elastic scattering power, though at the cost of a lower resolution (about 10 \AA{}), while the higher energy would be required for reaching a few \AA{}ngstr\"om resolution.

It is important to note that for our model and sample, the photon energy does not affect the qualitative ionization dynamics.
Increasing the photon energy has two major effects: the cross section of the photoionization is rapidly decreasing, and escaping photoelectrons have a higher kinetic energy, thus ionizing fewer atoms on their way out.
The latter ionization process is not dominant for our sample, and the former process merely leads to an effective rescaling of the fluence.
For light atoms (C, N, O) considered in this work the photoionization cross section at 3.1 keV is two orders of magnitude larger than at 12.4 keV.
Consequently, the ionization dynamics at the photon energy of 3.1 keV and fluence $F$ are similar to dynamics observed at 12.4 keV and a fluence $10^{2} \cdot F$.

All diffraction patterns were simulated in kinematic approximation using mode decomposition described in Ref. \cite{Ulf}.

\subsection{Ionization dynamics}

To describe the contribution of a specific ionization process $\alpha$, we first introduce several quantitative measures.
We define the probability $R^\alpha_i(t) dt$ that an atom of the type $i$, undergoes a state change due to a process $\alpha$ during a time interval $[t, t+dt]$ as
\begin{equation}
R_i^\alpha(t) = \sum_\xi \sum_{\eta \neq \xi} R^\alpha_{\xi \eta; i}(t) p_{\eta; i}(t)
\end{equation}
Since all state changes in our model lead to the removal of a single electron, $R_i^\alpha(t)$ is also the rate of ionization.

To get a global measure of the ionization process, we can integrate these ionization rates over time to get the number of ionized electrons due to process $\alpha$, $\delta_i^\alpha(t) = \int_{-\infty}^t R_i^\alpha(\tau) d\tau$.
Normalizing this quantity to the number of electrons of the neutral atom, $Z_i$, and integrating with the normalized pulse shape $J(t)$, gives the average degree of ionization,
\begin{gather}
\label{theory::indiv_doi}
\Delta_i^\alpha = \frac{1}{I_0}\int_{-\infty}^\infty J(t) \frac{\delta_i^\alpha(t)}{Z_i} dt \ .
\end{gather}
The average degree of ionization is a quantity between 0 and 1 that determines how many electrons are lost due to process $\alpha$.
The weighting with the pulse shape guarantees that ionization is counted only during the time of the pulse propagation through the sample.
For example, if Auger ionization starts after the pulse has propagated through the sample, it will yield insignificant contribution to $\Delta_i^\alpha$, though the degree of ionization can be substantial.

For our analysis we also introduced quantities averaged over all atoms
\begin{align}
\label{eq::ionization_rate}
\overline{R}^\alpha(t) &= \sum_i^\text{C,N,O} w_i R^\alpha_i(t), \\
\label{eq::ionization_degree}
\overline{\Delta}^\alpha &= \sum_i^\text{C,N,O} w_i \Delta_i^\alpha, \\
\label{eq::scattering_power}
\overline{n}_b &= 1 - \sum_\alpha \overline{\Delta}^\alpha
\end{align}
with the weights $w_\text{C}, w_\text{N}, w_\text{O}$ of 3/5, 1/5 and 1/5, corresponding to contribution of C, N, and O, respectively.
The quantity $\overline{n}_b$ can be interpreted as the normalized average number of electrons bound to an atom during the pulse propagation, with $\overline{n}_b=1$ denoting an undamaged atom.

\begin{figure}
	\includegraphics[width=\linewidth]{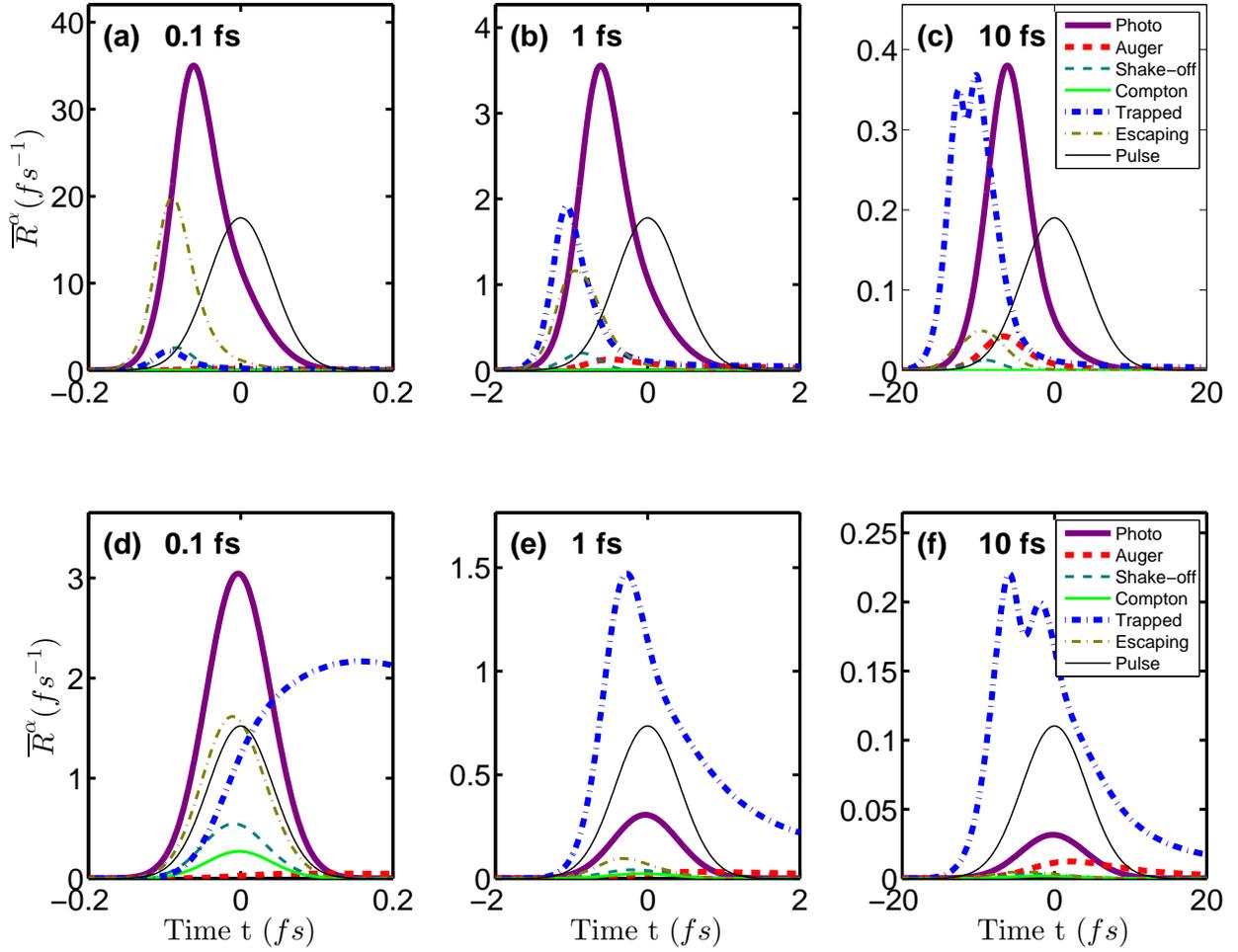}
	\caption{\label{fig::rates}
		(Color online) Average ionization rates $\overline{R}^\alpha(t)$ of different radiation processes for photon energies of 3.1 keV (a-c) and 12.4 keV (d-f).
		The pulse durations are 0.1 fs (a,d), 1 fs (b,e), and 10 fs (c,f).
		The pulse fluence is $10^{14}$ ph/$\mu$m$^2$ in all cases.
		The black thin solid line shows the pulse shape.
	}
	\newpage
\end{figure}

\begin{figure}
	\includegraphics[width=\linewidth]{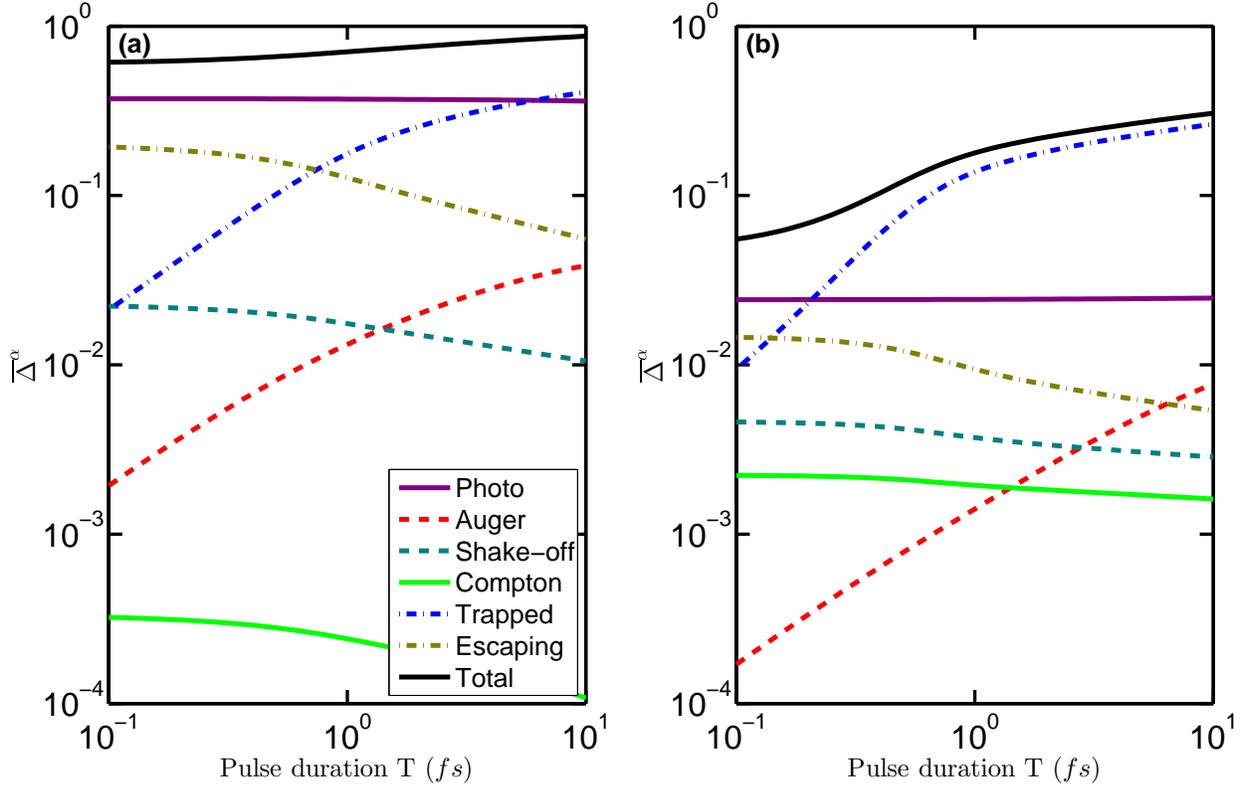}
	\caption{\label{fig::average_ionization}
		(Color online) Average degree of ionization $\overline{\Delta}^\alpha$ for different ionization processes and pulse durations for photon energies of 3.1 keV (a) and 12.4 keV (b).
		The pulse fluence is $10^{14}$ photons/$\mu$m$^2$ in both cases.
	}
	\newpage
\end{figure}

The time-dependent rates $\overline{R}^\alpha(t)$ for different ionization processes and average degrees of ionization $\overline{\Delta}^\alpha$ are presented in Fig.~\ref{fig::rates} and Fig.~\ref{fig::average_ionization}, respectively.
In both cases, results are shown for the two photon energies of 3.1 keV and 12.4 keV and the same fluence of 10$^{14}$ photons/$\mu$m$^2$.
Since the photoionization cross section drops by two orders of magnitude at the higher photon energy, this fluence corresponds to two different ionization regimes.
At 3.1 keV, most of electrons are removed from their atoms, so we call this the \emph{strong ionization regime}.
In contrast, at 12.4 keV about half of the electrons remain bound even at the end of the longest pulse, therefore we call this the \emph{weak ionization regime}.

From the data presented in Fig.~\ref{fig::rates} and Fig.~\ref{fig::average_ionization} we can conclude that:
\begin{enumerate}
	\item Only photoionization and impact ionization by trapped and escaping electrons contibute substantially to the direct ionization of the atoms.
	\item The net effect of photoionization is independent of the pulse duration (Fig.~\ref{fig::average_ionization}).
	Since the rate is a time derivative of the number of ionized electrons, the photoionization rate is inversely proportional to the pulse duration (Fig.~\ref{fig::rates}).
    \item As expected, for short pulses of 0.1 fs Auger process is reduced by two orders of magnitude in comparison to 10 fs pulses (see Fig.~\ref{fig::average_ionization}).
	\item Impact ionization from trapped electrons is the dominant ionization process at rather long pulse durations of 10 fs.
	If the pulse duration is reduced, this process is suppressed for the strong ionization regime (Fig.~\ref{fig::rates}(a,b)), and delayed to the end of the pulse for the weak ionization regime (Fig.~\ref{fig::rates}(d)).
	In both cases, the average degree of ionization from trapped electron ionization decreases with decreasing pulse duration (Fig.~\ref{fig::average_ionization}).
	While this decrease is particularly significant for sub-fs pulses, we point out that the total degree of ionization (see black curve in Fig.~\ref{fig::average_ionization}) already goes down by half if we reduce the pulse duration from 10 fs to 1 fs.
	\item With decreasing pulse duration, impact ionization by escaping photoelectrons becomes a relevant process (Fig.~\ref{fig::average_ionization}, \ref{fig::rates}).
	This process becomes especially important for sub-fs pulses, and it substitutes the ionization from trapped electrons in the strong ionization regime (Fig.~\ref{fig::rates}(a,b)).
\end{enumerate}

These findings can be explained from basic considerations.
Photoionization in the x-ray energy range is for all practical purposes an instantaneous process that only depends on the pulse fluence.
Hence, any reduction of the pulse length with a constant fluence leads to a corresponding increase in the photoionization rate without changing the ionization dynamics.
Note that the cross section for photoionization of valence electrons is an order of magnitude smaller than that of core electrons.
This effect causes the apparent shift between the photoionization rate and the pulse shape in Fig.~\ref{fig::rates}(a-c), where all core electrons are ionized at the onset of the pulse.
While this difference in cross sections can in principle be used to create hollow atoms \cite{Sang-Kil}, this does not play a role at these particular fluences.

The impact ionization by trapped electrons is hindered by three factors:
First, the cross section decreases by about a factor of three for each additional charge of the atom, hence, impact ionization becomes a slow process for highly charged atoms.
Second, as the atoms are ionized, the binding energy of the valence electrons increases rapidly.
Finally, the energy of the trapped electrons is replenished only by the Auger process.
These have a typical lifetime of several fs that increases further if there are fewer valence electrons to fill the core holes.
A decrease of the pulse duration therefore allows to outrun the Auger decay, which makes the impact ionization by trapped electrons as negligible.

We note also that the double-peak form of the trapped electron ionization rate in Fig.~\ref{fig::rates}(c,f) arises from an interplay of these factors.
The ionization rate drops initially because the trapped electrons cannot supply sufficient energy to ionize further atoms.
At that point, Auger decay sets in, leading to a second maximum, after which the sample becomes so strongly ionized that the impact ionization becomes inefficient.

For our sample consisting of light atoms, and for the considered photon energies, ionization from escaping photoelectrons is not as efficient, because the impact ionization cross sections drop rapidly with increasing electron kinetic energy.
At most about every second photoelectron ionizes an atom on its way out of the sample.
For long pulses of 10 fs (Fig.~\ref{fig::rates}(c,f)), the atoms have already been strongly ionized by the trapped electrons when the maximum of photoelectrons are produced.
However, for sub-fs pulses, the first photoelectrons encounter a sample of neutral atoms, making subsequent impact ionization more likely.
Hence, this process plays a role only for the shortest pulse durations 0.1 fs.
In the strong ionization regime it also appears only at the onset of the pulse (Fig.\ref{fig::rates}(a)).

\begin{figure}
	\includegraphics[width=\linewidth]{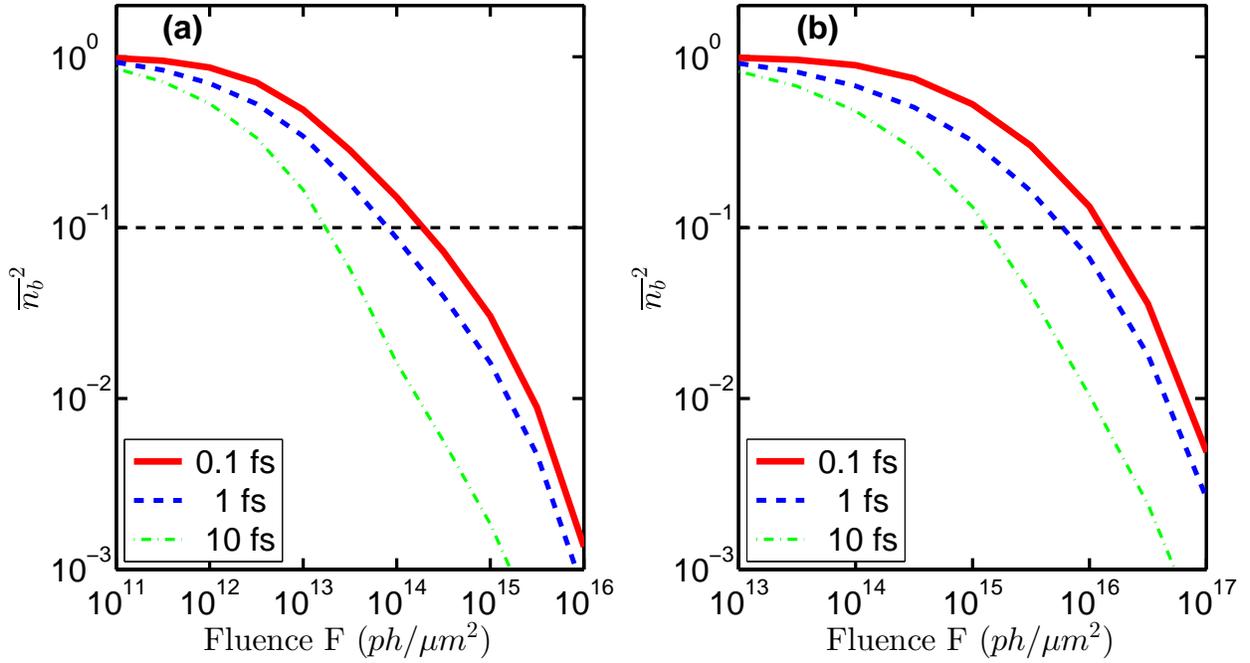}
	\caption{\label{fig::bound_electrons}
		(Color online) Square of the average normalized number of bound electrons, $\overline{n}_b^2$, for different pulse durations and fluences as a figure of merit of the samples scattering power.
		Photon energies are 3.1 keV (a) and 12.4 keV (b).
		The horizontal black dashed line corresponds to a cutoff of 10\% of the scattering power of the undamaged sample.
	}
	\newpage
\end{figure}

As a rough measure of the resulting scattering power of the sample, we can consider the square of the average number of bound electrons, $\overline{n}_b^2$, Eq.~\eqref{eq::scattering_power}.
The results for both photon energies and different pulse durations and fluences are shown in Fig.~\ref{fig::bound_electrons}.
The horizontal dotted line shows a cutoff where the sample retains approximately 10\% of its scattering power.
For 3.1 keV (Fig.~\ref{fig::bound_electrons}(a)) this cutoff is crossed at fluences from $10^{13}$ to $10^{14}$ photons/$\mu$m$^2$ for pulse durations from 10 fs to 0.1 fs, respectively.
The same behavior is observed for 12.4 keV photon energy (Fig.~\ref{fig::bound_electrons}(b)) where the cutoff is reached for fluences in the range from $10^{15}$ to $10^{16}$ photons/$\mu$m$^2$.
Altogether, we find that reducing the pulse duration from 10 fs to 1 fs significantly reduces the electronic radiation damage.
A further reduction to 0.1 fs yields another, but noticeably smaller reduction.

\subsection{Elastic and inelastic scattering}

As discussed in Section \ref{sec::theory::scattering}, the final signal at the detector has three contributions: the elastically scattered coherent signal $I_\text{W}(\mathbf{q})$, incoherent background $I_\text{B}(q)$, and Compton background $I_\text{Compton}(q)$.
Only the coherent signal $I_\text{W}(\mathbf{q})$ carries information about the particle internal structure.
It is therefore important to understand how electronic damage influences the coherent signal and background contributions.

\begin{figure}
	\includegraphics[width=\linewidth]{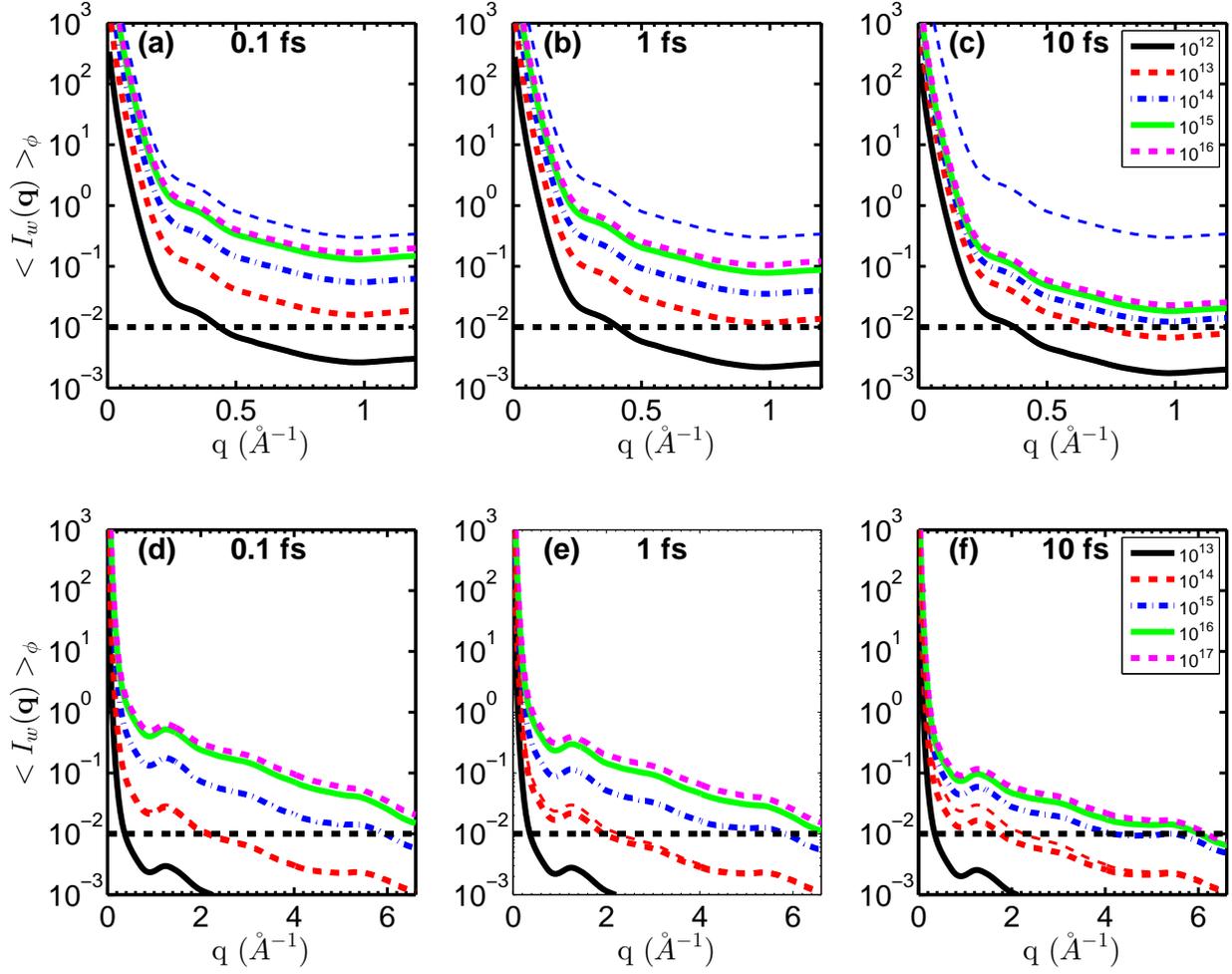}
	\caption{\label{fig::angular_intensity}
		(Color online) Angular averaged coherent signal $\langle I_\text{W}(\textbf{\textbf{q}})\rangle_{\phi}$ for photon energies of 3.1 keV (a-c) and 12.4 keV (d-f) and different fluences, defined in $ph/\mu m^2$ in the insets.
		Pulse durations are $0.1$ fs (a,d), $1$ fs (b,e), and $10$ fs (c,f).
		The horizontal dashed black line corresponds to the requirement of $10^{-2}$ photons per resolution element.
		The curves have been smoothed with a Gaussian filter to remove high-frequency oscillations from the speckle pattern.
		Thin dashed lines correspond to undamaged sample and fluence $10^{14}\ ph/\mu m^2$.
	}
	\newpage
\end{figure}

The coherent signal as a function of the momentum transfer for different fluences and pulse durations is shown in Fig.~\ref{fig::angular_intensity}.
The intensity was calculated according to Eq.~\eqref{SII.3} and angularly averaged over all detector pixels of constant $|\mathbf{q}|$, giving the angular averaged intensity, $\langle I_\text{W}(\mathbf{q}) \rangle_\phi = (2\pi)^{-1} \int_0^{2\pi} I_\text{W}(q, \phi) d\phi$ per Shannon angle.

If we disregard the background contribution , we can define the maximum achievable resolution by requiring a minimum of $10^{-2}$ photons per Shannon angle for successful orientation \cite{orientation1, orientation2}.
In practice, this number may be higher due to artifacts and noise.
At 3.1 keV photon energy and a fluence of $10^{13}$ photons/$\mu$m$^2$, a pulse duration of 10 fs allows to achieve 8 \AA{} resolution, while 1 fs and 0.1 fs pulses allow to reach 4 \AA{}.
For 12.4 keV and a fluence of $10^{14}$ photons/$\mu$m$^2$, we can achieve about 3 \AA{} resolution; a further increase towards 1 \AA{} is only possible by increasing the fluence even further and at the same time having pulses of less then 1 fs duration.
Analysis of the results presented in Fig.~\ref{fig::angular_intensity} shows a substantial difference between the strong ionization regime Fig.~\ref{fig::angular_intensity}(a-c) and weak ionization regime Fig.~\ref{fig::angular_intensity}(d-f).
In the former case scattered intensities are substationally lower than intensities corresponding to an undamaged sample even at very short pulse durations of 0.1 fs (see Fig.~\ref{fig::angular_intensity}(a)) and in the latter case they are very close to each other.
Another important effect is the saturation of the scattered intensity.
At high fluences in the strong ionization regime an increase of the XFEL intensity by one order of magnitude does not lead to the same increase of the scattered intensity.

\begin{figure}
	\includegraphics[width=\linewidth]{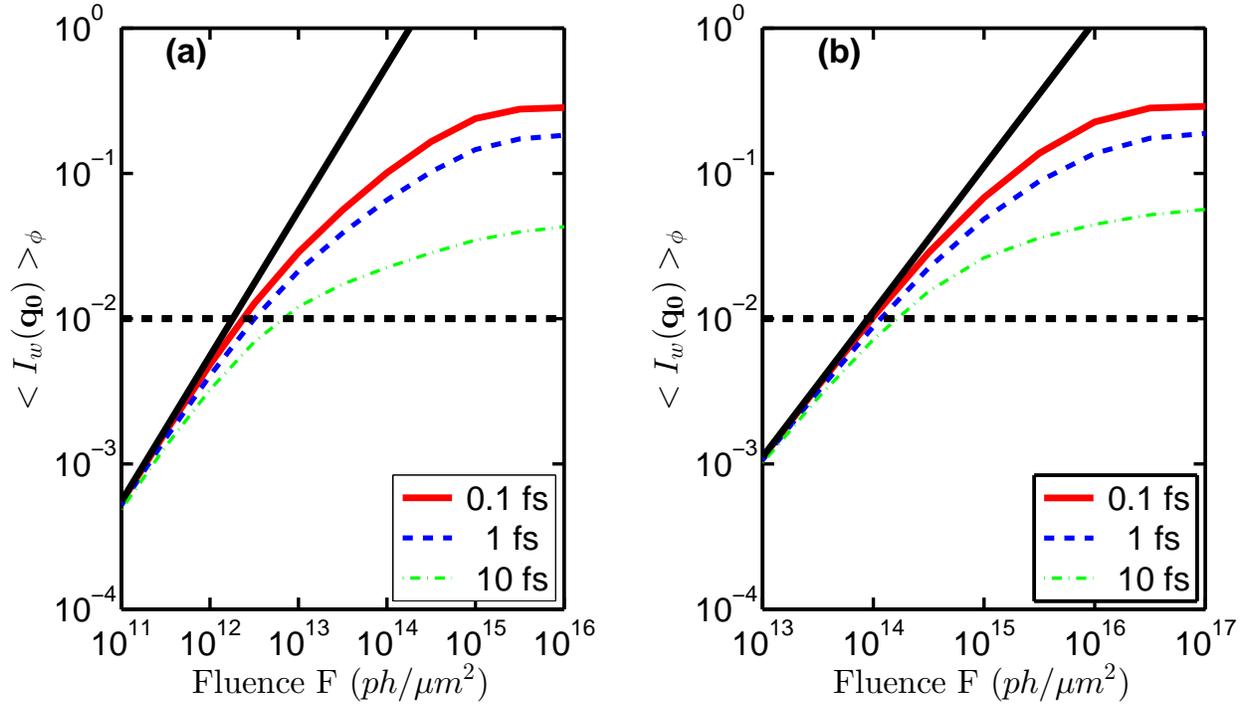}
	\caption{\label{fig::form_factor}
		(Color online) Angular averaged coherent signal $<I_W(q_0)>_{\phi}$ for different pulse durations and fluences, defined in $ph/\mu m^2$ in the insets.
		The dashed black line represents the lower limit of $10^{-2}$ photons per resolution element.
		The solid black line corresponds to undamaged sample.
		The quantities are evaluated at 10 \AA{} resolution for 3.1 keV photon energy (a) and 3 \AA{} resolution for 12.4 keV (b).
	}
	\newpage
\end{figure}

In the following, we consider a resolution of 10 \AA{} for 3.1 keV and 3 \AA{} for 12.4 keV, corresponding to $q_\text{0} \approx 0.6$ \AA$^{-1}$ and $q_\text{0} \approx 2$ \AA$^{-1}$, respectively.
Fig.~\ref{fig::form_factor} shows the averaged scattering intensities $\langle I_\text{W}(q_\text{0})\rangle_\phi$ at these momentum transfer values as a function of pulse duration and fluence.
%
%

For an undamaged sample, Fig.~\ref{fig::form_factor} shows a linear relationship between the incoming pulse fluence and scattered intensity.
However, due to ionization of the sample, this relationship breaks down at high XFEL intensities.
For 10 fs pulses, deviation from the linear scaling law starts at $10^{12}$ ($10^{14}$) photons/$\mu$m$^2$ for 3.1 (12.4) keV, and becomes significant at one order of magnitude higher fluence.
This is caused by a substantial decrease of the number of bound electrons that can scatter (see Fig.~\ref{fig::bound_electrons}).
Reducing the pulse duration to 1 fs increases the scattered intensity by reducing electronic radiation damage.
A further reduction to 0.1 fs gives another, but considerably smaller increase, in qualitative agreement with the radiation damage observables (Fig.~\ref{fig::bound_electrons}).

The results here put both lower and upper boundaries on acceptable fluences for imaging the test particle.
To achieve the required resolutions, a minimum fluence of $2\cdot 10^{12}$ ($10^{14}$) photons/$\mu$m$^2$ for 3.1 (12.4) keV is strictly required to get enough scattered signal.
At 10 fs pulse duration, however, an increase in fluence no longer translates into an increase in scattered intensity for intensities beyond $10^{13}$ ($10^{15}$) photons/$\mu$m$^2$.
An increase of the fluence by one order of magnitude increases the scattered signal only by a factor of two.
Decreasing the pulse duration to 1 fs already improves the scaling significantly, with another smaller gain when going to 0.1 fs pulses.
Still, even for the shortest 0.1 fs pulses, there is little advantage from increasing the fluence beyond $10^{14}$ (few $10^{15}$) photons/$\mu$m$^2$ for 3.1 (12.4) keV.

\bigskip

\begin{figure}
	\includegraphics[width=\linewidth]{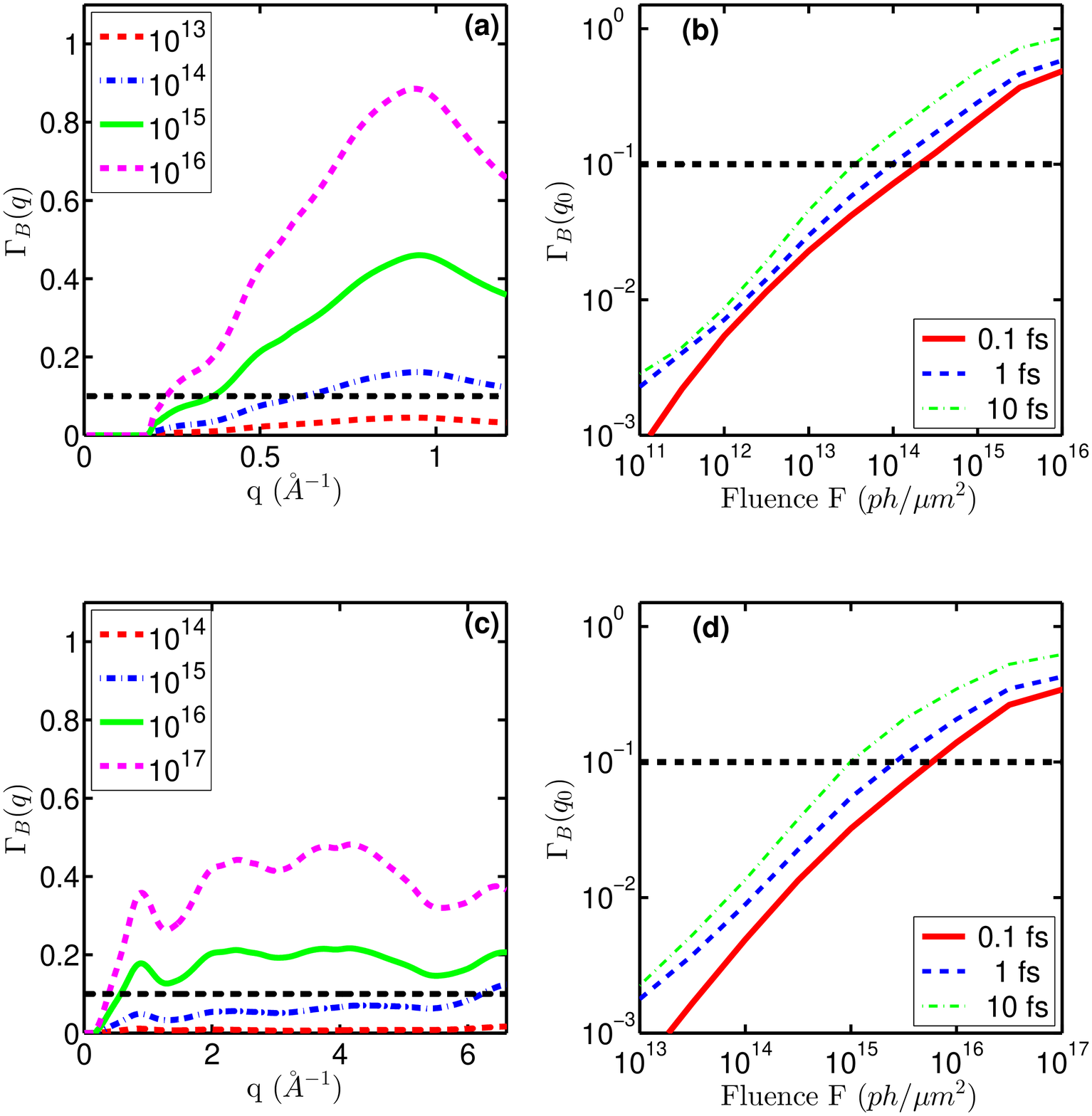}
	\caption{\label{fig::gamma}
		(Color online) (a,c) Relative contribution of the incoherent background, $\Gamma_B(q)$, as a function of momentum transfer for different fluences, defined in $ph/\mu m^2$ in the insets, photon energies of 3.1 keV (a) and 12.4 keV (c),  and pulse duration of 1 fs.
		(b,d) $\Gamma_B(q_0)$ as a function of pulse fluence for different pulse duration at a resolution of 10 \AA{} for 3.1 keV (b) and 3 \AA{} for 12.4 keV (d).
		The horizontal dashed black lines represent an upper acceptable limit of 10\% background.
	}
	\newpage
\end{figure}

As a simple measure of the background effects, we can use the ratio between the respective background contribution (incoherent signal $I_{\text{B}}(q)$ or Compton background $I_{\text{Compton}}(q)$) and the coherent signal,
\begin{gather}
\label{theory::backgroundContrib}
\Gamma_{\text{B/Compton}}(q) = \frac{I_{\text{B/Compton}}(q)}{\langle I_{\text{W}}(\mathbf{q}) \rangle_{\phi}}.
\end{gather}
This measure for the incoherent background $\Gamma_{\text{B}}(q)$ is shown in Fig.~\ref{fig::gamma}(a,c) for both photon energies and 1 fs pulse duration.
Note that the oscillations in these figures are caused by the speckle pattern of the diffraction image; the background $I_{\text{B}}(q)$  is a smooth function of $q$.
We considered a cutoff of 10\% shown as a horizontal dashed line, after which the background becomes a significant feature of the diffraction pattern and complicates the analysis, especially the orientation of the single-shot diffraction patterns.

The dependence of $\Gamma_B(q_\text{0})$ on the XFEL fluence and pulse duration is presented in Fig.~\ref{fig::gamma}(b,d).
The background rises continuously from negligible noise to the dominant contribution as the fluence increases.
For the highest fluences, $\Gamma_B(q_\text{0})$ shows saturation for all x-ray parameters, at values up to one.
A reduction of the pulse duration slightly reduces the background and the saturation value of $\Gamma_B(q_\text{0})$.
For 10 fs pulses, the cutoff is reached for fluences of $3\cdot 10^{13}$ ($10^{15}$) photons/$\mu$m$^2$ for 3.1 (12.4) keV x-rays.
Reducing the pulse duration further to 0.1 fs increases the allowed fluences to about $2\cdot 10^{14}$ ($6\cdot 10^{15}$) photons/$\mu$m$^2$.
Hence, the use of very short pulse durations is experimentally attractive to suppress this background contribution.
Note that the restrictions on the pulse fluence are similar to those from considering only the coherent scattering $I_\text{W}(\mathbf{q})$.

\bigskip

\begin{figure}
	\includegraphics[width=\linewidth]{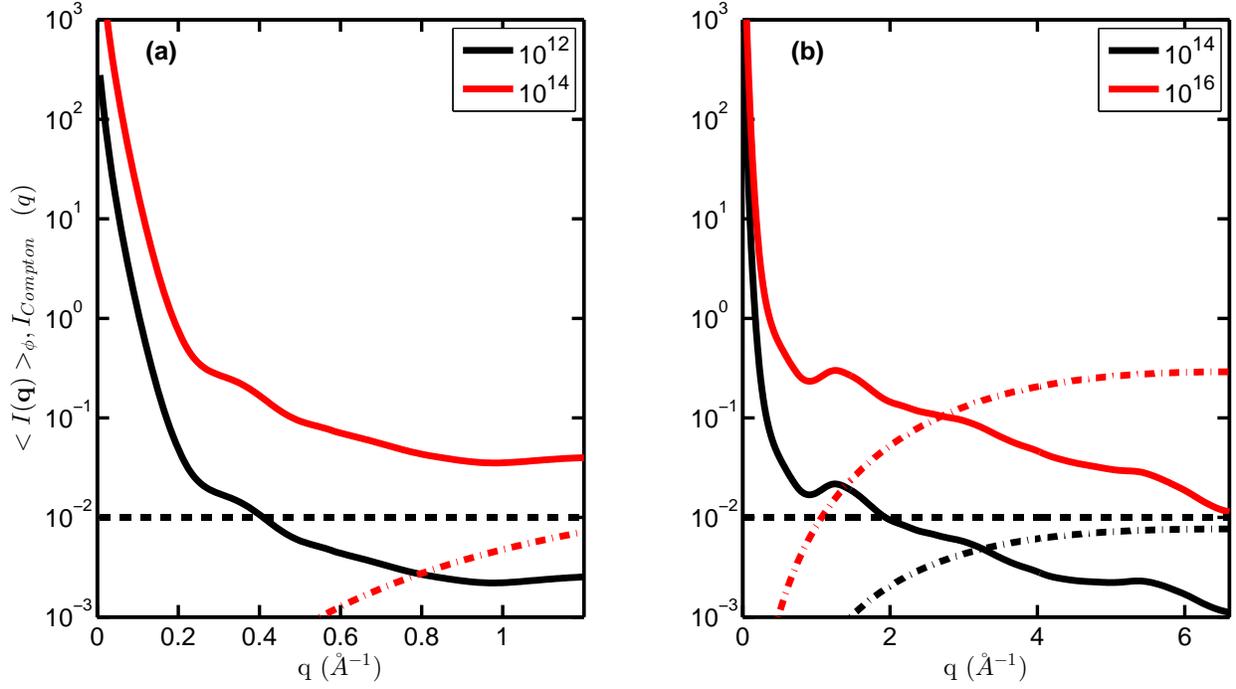}
	\caption{\label{fig::compton}
		(Color online) Comparison of the angular averaged coherent signal $\langle I_W(\mathbf{q})\rangle_{\phi}$ (solid lines) and Compton signal $I_\text{Compton}(q)$ (dashed lines) for photon energies of 3.1 keV (a) and 12.4 keV (b) and different fluences, defined in $ph/\mu m^2$ in the insets.
		The pulse duration is 1 fs in both cases.
		The horizontal dashed black line corresponds to the requirement of $10^{-2}$ photons per resolution element.
	}
	\newpage
\end{figure}

A comparison of the coherent signal and Compton scattering contribution is shown in Fig.~\ref{fig::compton} for pulses with 1 fs duration, different fluences and both photon energies.
Only the contribution from bound electrons was taken into account, the contribution from trapped and escaping electrons was neglected.
Hence, the presented results could be considered as a lower boundary.
As expected the Compton scattering becomes more important at high photon energy, and dominates the signal for high momentum transfers at $q \geq 3$ \AA$^{-1}$.
The Compton scattering is relatively weak for soft x-rays, never reaching the coherent signal.

\begin{figure}
	\includegraphics[width=\linewidth]{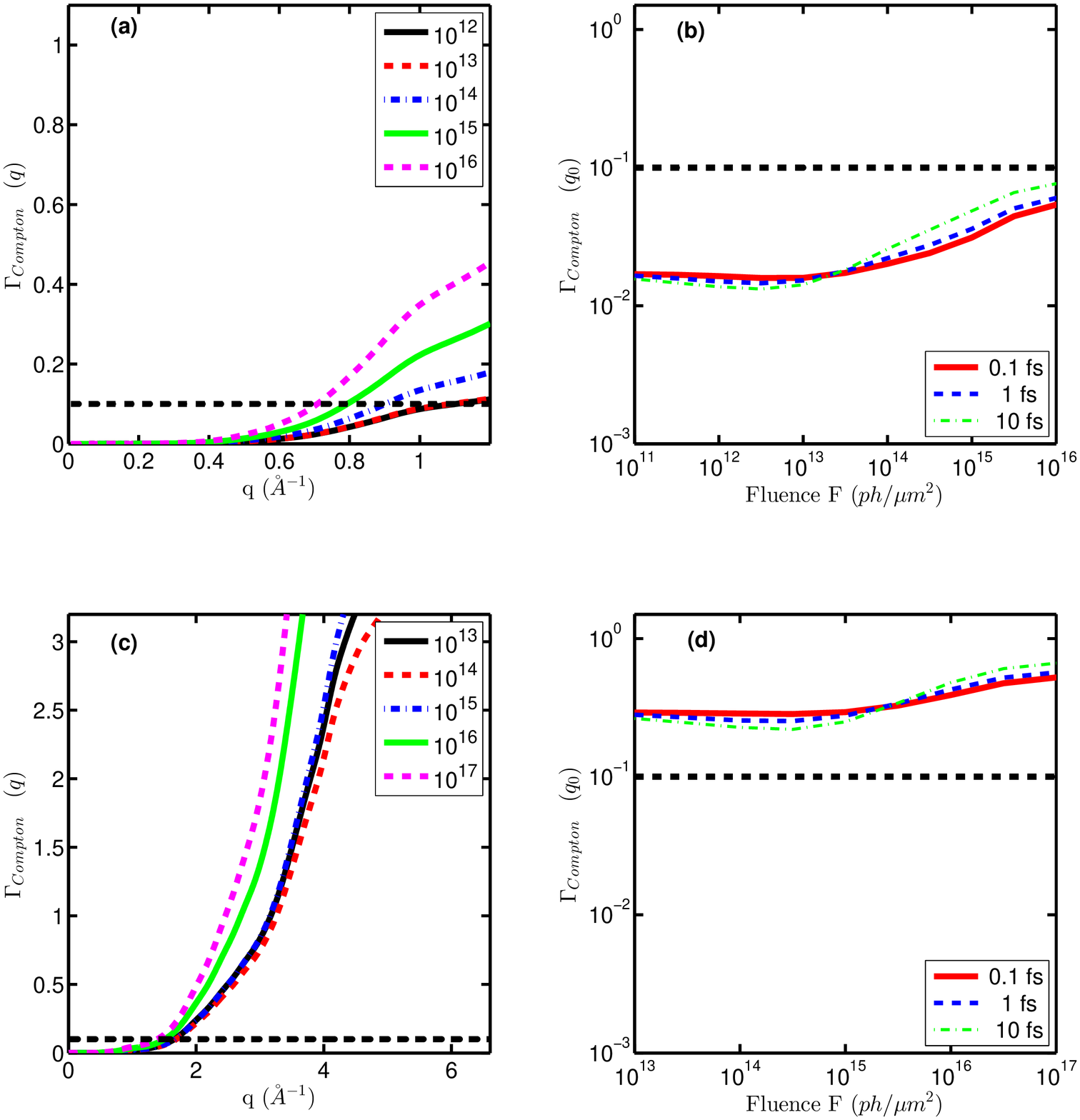}
	\caption{\label{fig::comptonback}
		(Color online) (a,c) Relative contribution of the Compton scattering, $\Gamma_\text{Compton}(q)$ as a function of momentum transfer for different fluences, defined in $ph/\mu m^2$ in the insets, photon energy of 3.1 keV (a) and 12.4 keV (c), and pulse duration of 1 fs.
		(b,d) $\Gamma_\text{Compton}(q_0)$ as a function of pulse fluence for different pulse durations at 10 \AA{} resolution for 3.1 keV (b) and 3 \AA{} resolution for 12.4 keV (d).
		The horizontal dashed black line represents an upper acceptable limit of 10\%.
	}
	\newpage
\end{figure}

The ratio $\Gamma_\text{Compton}$ (Eq.~\eqref{theory::backgroundContrib}) for different pulse parameters is shown in Fig.~\ref{fig::comptonback}.
We point out that in our simulations, the relative Compton background from bound electrons is larger than in \cite{SantraCompton}.
This increase is due to two factors.
The atoms in our simulations are stronger ionised on average due to inclusion of the electron impact ionization that leads to a stronger suppression of the coherent scattering.
Also, the explicit inclusion of nitrogen/oxygen atoms with more valence electrons (3 and 4 respectively versus 2 for carbon) increases the Compton scattering contribution, since this process dominantly occurs on weakly-bound electrons.

We found that the limit of 10\% background is always surpassed for the momentum transfer values larger than $q \geq 1$ \AA$^{-1}$ ($ q \geq 1.5$ \AA$^{-1}$) and photon energies of 3.1 keV (12.4 keV) (see Fig.~\ref{fig::comptonback}(a,c)).
We also observed that the dependence of the Compton background at a constant momentum transfer value $q_0$ on the pulse parameters is rather weak (see Fig.~\ref{fig::comptonback}(b,d)).
The Compton contribution practically does not depend on the pulse duration and $\Gamma_\text{Compton}(q_0)$ increases only by a factor of two to three for fluences above $F \geq 10^{13} (10^{15})$ photons/$\mu$m$^2$ at 3.1 (12.4) keV photon energy, respectively.
%
Effectively, considering the maximum acceptable background level to be 10\%, the Compton scattering limits the achievable resolution to approximately 6 \AA{} (4 \AA{}) for the soft (hard) x-rays.

Our simulations show that the Compton scattering gives a substantial contribution in the hard x-ray scattering conditions and less important in the soft x-ray range.
Without a proper treatment of this background or use of energy-resolved detectors, few \AA{}ngstr\"o{}m resolution limit will be difficult to reach for small non-crystalline particles.

\section{Summary and conclusions}
\label{sec::summary}

In summary, we have extended our previous approach \cite{Ulf} on ionization dynamics of biological samples to incorporate shake-off ionization and Compton scattering.
We studied the ionization dynamics and the scattered signal for ultrashort XFEL pulses from 0.1 fs to 10 fs.
We used an adenovirus shell as a test sample, and considered soft(hard) x-ray pulses with 3.1(12.4) keV photon energy and 10(3) \AA{} target resolution, respectively.

By introducing appropriate measures, we quantified the contribution of the single ionization mechanisms to the electronic radiation damage.
In particular, we found that with sufficiently short pulses (on the order of 1 fs and less) it is possible to outrun the ionization from the trapped electron gas and therefore to reduce the electronic damage significantly.
Our simulations show that it is \emph{not} necessary to use sub-fs pulses; a considerable damage reduction is already realized for pulse durations of 1 fs.

To translate this into a useful fluence limit, we also analyzed the scattered intensity, which has three contributions.
These are a coherent signal that contains all the structural information, an incoherent background that is due to statistical fluctuations of the form factors of individual atoms, and the Compton (inelastic) background.
For a given sample, they put different boundaries on the XFEL fluence and achievable resolution in different ways.
If the XFEL fluence is too large, most of electrons are striped from the atoms and the sample does not scatter anymore.
In this case scenario the coherent signal does not increase any more with the increasing XFEL pulse fluence.

We found that for a given XFEL fluence the use of shorter pulses increases the coherent signal and reduces the incoherent background.
In concord with the radiation damage measures, we found a large improvement already by reducing the pulse duration from 10 fs to 1 fs.
Even with the shortest pulses of 0.1 fs, however, there is an upper limit of about $10^{14}$ (few $10^{15}$) photons/$\mu$m$^2$ for 3.1 (12.4) keV photon energy.
Also at these fluences the incoherent background becomes relevant, the signal-to-noise ratio drops down, and recovery of a structural information will become a challenging task.

We also studied the Compton background, and found that it provides a strong contribution to the scattering signal at a few {\AA}ngstr\"{o}m resolution.
For our test sample, we found that Compton scattering limits the available resolution to about 6 \AA{} (4 \AA) for a photon energy of 3.1 (12.4) keV.
This background is virtually independent of the pulse parameters, and cannot be suppressed by the use of short pulses or small fluences.
To overcome this problem a special study of the Compton contribution and, possibly, dedicated energy resolution detectors will be necessary in the future.

Alltogether, our analysis shows that SPI experiments are still challenging, especially for small biological samples of 30 nm size and below.
To reach subnanometer resolution we suggest to use pulses of about 1 fs and the XFEL fluence that is below high ionization regime.
In order to obtain access to high resolution a substantial amount of diffraction patterns should be accumulated at these XFEL conditions.
Our results show that it is not advisable to go to high ionization regime of XFEL operation since ionization dynamics prevents substantial increase of the scattered intensity with the raise of the XFEL power.



\begin{acknowledgments}
The support of the project and fruitful discussions with E. Weckert, as well as careful reading of the manuscript by T. Laarman are greatly acknowledged.
Ulf Lorenz thanks the Deutsche Forschungsgemeinschaft (DFG) for financial support through Project No. Sa 547/9.
Nikolay Kabachnik acknowledges financial support by European XFEL (Hamburg) and from the programme "Physics with Accelerators and Reactors in West Europe" of the Russian Ministry of Education and Science. 
\end{acknowledgments}

\appendix

\section{Elastic scattering of extremely short pulses}
\label{appendix::scattering}

Here we present the general formalism of elastic scattering of ultrashort FEL pulses on a single particle in kinematical approximation.



We consider the amplitude of the incoming x-ray wave in the form

\begin{equation}
\widetilde{E}_{in}({\bf r},t) = E_{in}({\bf r},t)\exp(i{\bf k_i} \cdot {\bf r}-i\omega t).
\label{C1}
\end{equation}
Here $E_{in}({\bf r},t)$ is the slowly varying amplitude, $\omega$ and ${\bf k_i}=(2\pi/\lambda) {\bf n_i}$ are the average frequency and momentum of the incoming x-ray wave with the average wavelength $\lambda$. The incident direction of the wave is defined by the vector ${\bf n_i}$.

In the frame of the first Born approximation the instantaneous amplitude of the outgoing wave at the time $t$ at the detector position coordinate ${\bf u}$ for a narrow bandwidth light can be presented as \cite{BornWolf}

\begin{equation}
{E}_{out}({\bf u},t) = \frac{1}{i\lambda} \int d {\bf r} \rho({\bf r}, t-\tau_{r})\frac{\widetilde{E}_{in}({\bf r},t-\tau_{r})}{R_{ur}},
\label{C2}
\end{equation}
where $\tau_{r}=R_{ur} / c$ is the time delay for the light propagating from the position {\bf r} in the sample to the position $\bf u$ in the detector and $R_{ur}$ is the distance between these two points (see Fig. A1).
Substituting now expression (\ref{C1}) into Eq. (\ref{C2}) we find for the amplitude of the scattered wave
\begin{equation}
{E}_{out}({\bf u},t) =  \frac{1}{i\lambda} \int d {\bf r} \rho({\bf r}, t-\tau_{r})\frac{{E}_{in}({\bf r},t-\tau_{r})}{R_{ur}}
e^{i{\bf k_i} \cdot {\bf r}-i\omega (t-\tau_r)}.
\label{C3}
\end{equation}
In the Fresnel limit we have for the distance $R_{ur}$ between the points {\bf r} and $\bf u$
\begin{equation}
R_{ur} = L- {\bf n_f \cdot r} + \frac{({\bf u} - {\bf r})^2}{2L},
\label{C4}
\end{equation}
where $L$ is a distance between the sample and the detector and vector $\bf n_f = {\bf k_f}/ |k|$ defines the direction of the outgoing wave.
Assuming far-field limit ($d^2/\lambda L << 1$, where $d$ is the size of the sample), as is typical in the case of the single particle imaging experiments and substituting (\ref{C4}) in Eq. (\ref{C3}) we find for the scattered field
%
%
%
\begin{eqnarray}
{E}_{out}({\bf u},t) & = & \frac{e^{ikL - i\omega t}}{i\lambda L} e^{iku^2/2L} A({\bf q}, t); \nonumber\\
A({\bf q}, t) & = & \int d {\bf r} \rho({\bf r}, t-\tau_{r}) {E}_{in}({\bf r},t-\tau_{r}) e^{-i{ \bf q}  \cdot {\bf r}},
\label{C6}
\end{eqnarray}
where the momentum transfer vector ${\bf q} = ({\bf k_f} - {\bf k_i})+ {\bf q_u}$ with ${\bf q_u} =k ({\bf u}/ L)$ is introduced.
This is the far-field expression for the instantaneous value of the scattered field at the detector position $\bf u$.

We write now the electron density of a single particle in a usual way as a sum of instantaneous electron densities $\rho_i({\bf r}, t)$ of each atom at the position ${\bf R}_i$
\begin{equation}
\rho({\bf r}, t) = \sum^N_{i=1} \rho_i({\bf r-R}_i, t).
\label{C7}
\end{equation}
We assume now that for the sufficiently short femtoseconds pulses considered here atomic positions do not change during the pulse propagation and all time dependencies are due to electronic changes in the individual atoms consisting the particle.
We also neglect here all cooperative effects.
Substituting expression (\ref{C7}) for the electron density into Eq. (\ref{C6}) and performing the change of variables we obtain for the amplitude of the scattered field {\footnote{We omit here and below all not important prefactors and we dropped the subscript index $\bf u$.}}
\begin{equation}
{A}({\bf q},t) =   \sum^N_{i=1} e^{-i{\bf q} \cdot {\bf R}_i}
\int d {\bf r'} \rho_i({\bf r'}, t-\tau_{R_i})
{E}_{in}({\bf R}_i + {\bf r'},t-\tau_{R_i}) e^{-i{ \bf q}  \cdot {\bf r'}}.
\label{C8}
\end{equation}
Assuming here that the incident x-ray field is uniform on the size of a single atom ${E}_{in}({\bf R}_i + {\bf r'},t-\tau_{R_i}) \approx {E}_{in}({\bf R}_i,t-\tau_{R_i})$ and introducing the time-dependent atomic form factors $f_i({\bf q}, t) = \int\rho_i({\bf r}, t)exp{(-i {\bf q} \cdot {\bf r})}d{\bf r}$ we obtain the following general expression for the scattered amplitude (\ref{C8})

%
%

\begin{equation}
{A}({\bf q},t) =   \sum^N_{i=1} f_i({\bf q}, t-\tau_{R_i})
{E}_{in}({\bf R}_i,t-\tau_{R_i}) e^{-i{ \bf q}  \cdot {\bf R}_i}.
\label{C9}
\end{equation}
This expression for the scattered amplitude from a single particle differs from the traditionally used kinematical expression by two important features. First, it contains the time delayed atomic form factors $f_i({\bf q}, t-\tau_{R_i})$ and, second, it contains the instantaneous time delayed incident field ${E}_{in}({\bf R}_i,t-\tau_{R_i})$  at each atomic position.


For a typical single particle imaging experiment diffraction pattern recorded on the detector will be given by the intensity of the wavefield defined by Eq. (\ref{C9}).
As soon as present detectors do not have femtosecond time resolution the measured signal will be, necessarily, integrated over the time of the pulse duration and will be given by
\begin{gather} \label{C10a}
\begin{align}
I({\bf q}) & =  \int I({\bf q},t) dt  =  \sum^N_{i,k=1} e^{-i{ \bf q}  \cdot ({\bf R}_k - {\bf R}_i)} \times  \notag \\
                 & \times \int dt f_i^{*}({\bf q}, t-\tau_{R_i}) f_k({\bf q}, t-\tau_{R_k})
{E}_{in}^{*}({\bf R}_i,t-\tau_{R_i}) {E}_{in}({\bf R}_k,t-\tau_{R_k}).
\end{align}
\end{gather}

As soon as this experiment will be repeated many times we would have to average results of these measurements.
This is equivalent to ensemble averaging of time-integrated intensity distribution (\ref{C10a})

\begin{gather} \label{C10}
\begin{align}
<I({\bf q})> & =  \int <I({\bf q},t)> dt  =  \sum^N_{i,k=1} e^{-i{ \bf q}  \cdot ({\bf R}_k - {\bf R}_i)} \times  \notag \\
                 & \times \int dt <f_i^{*}({\bf q}, t-\tau_{R_i}) f_k({\bf q}, t-\tau_{R_k})
{E}_{in}^{*}({\bf R}_i,t-\tau_{R_i}) {E}_{in}({\bf R}_k,t-\tau_{R_k})>.
\end{align}
\end{gather}

Assuming now that fluctuations of the incoming wavefield are statistically independent from the fluctuations of the electronic system  we can factorize averaging in Eq. (\ref{C10}) into a product of two terms


%
\begin{gather} \label{C12}
\begin{align}
<I({\bf q})> & =  \sum^N_{i,k=1} e^{-i{ \bf q}  \cdot ({\bf R}_k - {\bf R}_i)} \times \notag \\
               &  \times \int dt <f_i^{*}({\bf q}, t-\tau_{R_i}) f_k({\bf q}, t-\tau_{R_k})> \times \notag \\
                &  \times <{E}_{in}^{*}({\bf R}_i,t-\tau_{R_i}) {E}_{in}({\bf R}_k,t-\tau_{R_k})>.
\end{align}
\end{gather}

At that stage we introduce for the incoming x-ray field a mutual coherence function \cite{MandelWolf}
\begin{equation}
\Gamma_{in} ({\bf r_1, r_2 }; t_1, t_2) =  \left< E_{in}^*({\bf r_1 }; t-t_1) E_{in}({\bf r_2 }; t-t_2) \right>
\label{C13}
\end{equation}
and its normalized version the so-called complex degree of coherence

\begin{equation}
\gamma_{in} ({\bf r_1, r_2};t_1, t_2) = \Gamma_{in} ({\bf r_1, r_2};t_1, t_2)/\sqrt{J ({\bf r_1};t)}\sqrt{J ({\bf r_2};t)},
\label{C15}
\end{equation}
where the incident intensity is defined as
\begin{equation}
J ({\bf r};t) = \Gamma_{in} ({\bf r_1 = r_2 = r}; t_1=t_2) = \left< \left|{E}_{in}({\bf r},t)\right|^2 \right>.
\label{C14}
\end{equation}
%

%
%
Substituting these definitions in Eq. (\ref{C12}) we obtain
\begin{gather} \label{C16}
\begin{align}
<I({\bf q})> & =  \sum^N_{i,k=1} e^{-i{ \bf q}  \cdot ({\bf R}_k - {\bf R}_i)} \times \notag \\
                 &  \times \int dt <f_i^{*}({\bf q}, t-\tau_{R_i}) f_k({\bf q}, t-\tau_{R_k})> \times \notag \\
                 & \times \sqrt{J ({\bf R}_i;t)}\sqrt{J ({\bf R}_k;t)}
                 \gamma_{in} ({\bf R}_i, {\bf R}_k; t, \tau_{ik}) ,
\end{align}
\end{gather}
where $\tau_{ik}=\tau_{R_i}-\tau_{R_k}$.

This is very general expression for the averaged intensity in a single particle imaging experiments derived in kinematical approximation.
It takes into account degradation of contrast of the coherently scattered intensity from a single particle due to two effects.
First, due to time evolution of the electronic structure of each atom as a result of fast ionization while propagation of the femtosecond x-ray pulses.
Second, due to partial coherence (spatial and temporal) of the incoming radiation.
As it was demonstrated in a series of experiments \cite{Vartanyants2011, Singer2012} x-ray pulses from XFEL sources have a high degree of spatial coherence and limited temporal coherence that could, in principle, degrade the contrast of the scattered intensity at high resolution.

At the next stage we will assume that the incoming radiation is fully coherent spatially and temporally, we will neglect retardation effects, and consider that the intensity distribution is spatially uniform over the size of the sample.
This leads to the following expression for the scattered intensity
\begin{gather}
\begin{align}
<I({\bf q})> & =  \sum^N_{i,k=1} e^{-i{ \bf q}  \cdot ({\bf R}_k - {\bf R}_i)} \times \notag \\
                 &  \times \int dt J (t) <f_i^{*}({\bf q}, t) f_k({\bf q}, t)> \ ,
\end{align}
\label{C16a}
\end{gather}
This expression was used as a starting point for the analysis in our previous paper \cite{Ulf} and in the present work.

\section{Details on the rate calculation}
\label{appendix::rates}

The direct photoionization rate is calculated as
$\hat{\mathbf{R}}^{\text{photo}}(t) = \hat{\boldsymbol{\sigma}}^{\text{photo}} j(t),$
where $\hat{\boldsymbol{\sigma}}^{\text{photo}}$ is the photoionization cross section {\footnote{Note that in this section we omit the index $i$ for the atom type where it is not specified.}}.
Since in the considered energy range the cross section is a slowly varying function of energy we use the cross section at the central frequency of the x-ray pulse.
The photoionization cross sections $\hat{\boldsymbol{\sigma}}^{\text{photo}}$ as well as the Auger
rates $\hat{\mathbf{R}}^{\text{Auger}}$ were calculated within the HFS approximation \cite{HermanSkillman}; the explicit expressions
can be found, for example, in \cite{Sang-Kil}.

During photoionization of core electrons, especially at large photon energies, the screening potential which is felt by other
electrons changes abruptly which can lead to the further emission of one or more electrons in a so-called shake-off process \cite{Carlson}.
If the photoionization transition occurs from the state $\eta$ to the state $\zeta$, it can be accompanied by the shake-off transition to the state $\xi$ with the probability $P^{\xi, \text{shake}}_{\zeta\eta}$.
Then the rate of such transition is $P^{\xi,\text{shake}}_{\zeta\eta}R^{\text{photo}}_{\zeta\eta}(t)$ and the total rate of various shake-off transitions to the same state $\xi$ is
${R}^{\text{shake}}_{\xi\eta}(t) = \sum_{\zeta}P^{\xi,\text{shake}}_{\zeta\eta}R^{\text{photo}}_{\zeta\eta}(t)$.

The Compton ionization rates are defined as
$\hat{\mathbf{R}}^{\text{Compton}}(t) = \hat{\boldsymbol{\sigma}}^{\text{Compton}} J(t)$.
The Compton cross sections $\hat{\boldsymbol{\sigma}}^{\text{Compton}}$  were
calculated in impulse approximation \cite{Rodberg} as \cite{Hubbell01, Hubbell1980}
\begin{gather}\label{eq:Compton}
    \sigma^{\text{Compton}}_{m;\xi} = \int\frac{d\sigma_{KN}(\theta)}
    {d\Omega}S_{m;\xi}(q)d\Omega \ .
\end{gather}
Here $d\sigma_{KN}(\theta)/d\Omega$ is the Klein-Nishina differential cross section, $\theta$ is the scattering angle, and $S_{m;\xi}(q)$ is the incoherent scattering function for electrons at the $m$-th shell for an atom in initial state $\xi$.
The cross section $\sigma^{\text{Compton}}_{m;\xi}$ describes the Compton process in which an electron is ejected from the m-th shell of an atom in the state $\xi$.
For low photon energies
$E_\gamma$ ($\epsilon = E_\gamma/m_ec^2 \ll 1$)
the Klein-Nishina differential cross section within a good approximation can be simplified to \cite{Jackson}
\begin{gather}\label{eq:Klein-Nishina}
    \frac{d\sigma_{KN}(\theta)}{d\Omega} \approx r^2_e[1+\epsilon (1-\cos \theta)]^{-2}P(\theta) \ ,
\end{gather}
where $r_e$ is the classical electron radius and $P(\theta)$ is the polarization coefficient.
In our simulations for photon energies up to 12 keV we used this approximation for Klein-Nishina cross section.

The incoherent scattering function for electrons at the $m$-th shell $S_{m;\xi}(q)$ was calculated in the frame of HFS approximation as \cite{Hubbell01}

\begin{gather}\label{eq:S(q)}
    S_{m;\xi}(q) = Z_{m;\xi} - \sum_{i=1}^{Z_{m;\xi}}|f^{i}(q)|^2 \ ,
\end{gather}
where $f^{i}(q)$ is the form factor of $i$-th electron on the $m$-th shell, and $Z_{m;\xi}$ the number of $m$-shell electrons in state $\xi$.
Here we neglect effects that may forbid excitation of an electron from one orbit to another due to Pauli exclusion principle.

It is well established (see for example Refs. \cite{Hau-Riege04,Ulf}) that secondary ionization can significantly change the ionization behavior in the biological particle.
In our model we treat the contribution of the secondary ionization similar to our previous work \cite{Ulf} (see also \cite{Hau-Riege04})
and neglect the details of the trapping process as a function of the particle charge and electron dynamics.
However, our model reproduce well most of the important features of the secondary ionization discussed in Ref. \cite{Hau-Riege04}.

We assume that photoelectrons originate in the center of the biological particle and during their escape produce secondary electrons through impact ionization. The biological particle is considered to be in the form of homogeneous sphere with the radius $R$ and volume $V$.
The rate of secondary ionization produced by the escaping high-energy photoelectrons is calculated according to \cite{Hau-Riege04, Ulf}
\begin{gather}
    \hat{\mathbf{R}}^{\text{escape}}(t) = \hat{\boldsymbol{\sigma}}^{\text{impact}}
        \frac{R}{V} \ \frac{\mathrm{d}N_{\text{photo}}}{\mathrm{d}t}
    \ ,
\end{gather}
where $\text{d}N_{\text{photo}} / \text{d}t$ is the production rate of photoelectrons.
The impact ionization cross sections
$\hat{\boldsymbol{\sigma}}^{\text{impact}}$ were calculated using the
binary-encounter Bethe model \cite{NIST} with electron orbital
parameters obtained from the HFS calculations. When a secondary
electron is produced, its kinetic energy was set to a constant value
$E_0=25$ eV \cite{Hau-Riege04}.

In our model Auger, shake-off, Compton as well as all secondary electrons produced by
electron impact ionization were supposed to be trapped by the positively charged particle and thermalized
instantaneously into a homogeneous electron gas with a
Maxwell-Boltzmann distribution of velocities.
The temperature $T$ of the gas was determined by the average kinetic energy of the
trapped electrons as $E_{trap} = (3/2) kT$, where $k$ is a Boltzmann constant.
The rate of collisional ionization produced by these trapped electrons was determined by
\begin{gather*}
    \hat{\mathbf{R}}^{\text{trap}}(t) = \erw{\hat{\boldsymbol{\sigma}}^{\text{impact}}(v) v}_T \ n_{\text{trap}}(t)
    \ ,
\end{gather*}
where we specifically introduced dependence on electron velocity $v$ in impact ionization cross section $\hat{\boldsymbol{\sigma}}^{\text{impact}}(v)$.
Here, $n_{\text{trap}}(t)$ is the time-dependent density of trapped electrons and $\erw{}_T$ is thermodynamic average with the appropriate Boltzmann factors.

The time evolution of the trapped electron density $n_{\text{trap}}(t)$ and total kinetic energy $E_{\text{trap}}(t)$ of the trapped
electron gas were calculated from the set of equations
\begin{gather}
\begin{align}
    \frac{\mathrm{d}n_{\text{trap}}}{\mathrm{d}t} &= \sum_{i} \sum_{\xi\eta}
        \biggl( R^{\text{Auger}}_{\xi\eta; i} + R^{\text{escape}}_{\xi\eta;i}(t)
        + R^{\text{trap}}_{\xi\eta;i}(t) + R^{\text{shake}}_{\xi\eta;i}(t)
        + R^{\text{Compton}}_{\xi\eta;i}(t) \biggr)p_{\eta;i}(t) \varrho_{i},\\
    \frac{\text{d}E_{\text{trap}}}{\text{d}t} &=
         \sum_{i} \sum_{\xi\eta} \biggl(
        R^{\text{Auger}}_{\xi\eta; i} E_{\xi\eta;i}^{\text{Auger}}
        + R^{\text{escape}}_{\xi\eta;,i}(t) E_0
        + R^{\text{shake}}_{\xi\eta;i}(t) E_{\xi\eta;i}^{\text{shake}} \notag \\
        & + R^{\text{Compton}}_{\xi\eta;i}(t) E_{\xi\eta;i}^{\text{Compton}}
        - R^{\text{trap}}_{\xi\eta;i}(t) E_{\xi\eta;i}^{\text{coll}} \biggr)
        p_{\eta;i}(t) \varrho_{i} V
    \ .
\end{align}
\end{gather}
Here the index $i$ denotes the atom type (carbon, nitrogen etc.)
with  the corresponding atom density $\varrho_{i}$.
The energy $E_{\xi\eta;i}^{\text{Auger}}$ is the energy of the
released Auger electrons, which is assumed to be approximately a
constant $E_{\xi\eta;i}^{\text{Auger}} \approx
E_i^{\text{Auger}}$ characteristic for a particular atom.
The energy $E_{\xi\eta;\alpha}^{\text{shake}}$ of the released
shake-off electrons is assumed to be about $1.8$ times the
binding energy of the electron \cite{Carlson}.
The energy of the electrons ionized by Compton scattering $E_{\xi\eta;i}^{\text{Compton}} \approx E_{i}^{\text{Compton}}$ is assumed
to be constant for a particular atom type.
The energy $E_{\xi\eta;\alpha}^{\text{coll}}$ is the binding energy of the electron that is released through impact ionization, and $E_0=25$ $eV$ is the energy assigned to secondary electrons produced by photoelectrons.

%
%

\bibliography{references}

\newpage

\begin{table}
	\caption{\label{tab::parameters}Parameters used in the simulations.}
	\begin{tabular}{l|c|c}
		photon energy (keV)                               & 3.1           & 12.4 \\
		pulse duration (fs)                               & 0.1 - 10      & 0.1 - 10 \\
		sample-detector distance (mm)                     & 100           & 100  \\
		detector size (mm)                                & 200           & 400  \\
		number of pixels                                  & 320 x 320     & 1536 x 1536 \\
		pixels per speckle                                & 2.4           & 1.4 \\
		Achievable resolution at the detector edge (\AA{})  & 5.2           & 0.95 \AA{}
	\end{tabular}
\end{table}


\end{document}